\documentclass[10pt,conference]{IEEEtran}
\IEEEoverridecommandlockouts
\usepackage{cite}
\usepackage{amsmath,amssymb,amsfonts}
\usepackage{algorithmic}
\usepackage{graphicx}

\usepackage[export]{adjustbox}

\usepackage{textcomp}
\usepackage{enumitem}
\usepackage{multirow}
\usepackage{amsmath}    
\usepackage{hyperref}   
\usepackage{diagbox}    

\usepackage{listings}

\usepackage[table,xcdraw]{xcolor}

\definecolor{codegreen}{rgb}{0,0.6,0}
\definecolor{codegray}{rgb}{0.5,0.5,0.5}
\definecolor{codepurple}{rgb}{0.58,0,0.82}
\definecolor{backcolour}{rgb}{0.95,0.95,0.95}

\definecolor{pastelyellow}{rgb}{0.99, 0.99, 0.59}
\definecolor{lightgreen}{rgb}{0.56, 0.93, 0.56}
\definecolor{lightcoral}{rgb}{0.94, 0.5, 0.5}

\lstdefinestyle{mystyle}{
  backgroundcolor=\color{backcolour},   commentstyle=\color{codegreen},
  keywordstyle=\color{magenta},
  numberstyle=\tiny\color{codegray},
  stringstyle=\color{codepurple},
  basicstyle=\ttfamily\scriptsize,
  breakatwhitespace=false,         
  breaklines=true,                 
  captionpos=b,                    
  keepspaces=true,                 
  numbers=left,                    
  numbersep=2pt,                  
  showspaces=false,                
  showstringspaces=false,
  showtabs=false,                  
  tabsize=2,
  frame=lines
}
\definecolor{light-gray}{gray}{0.95} 
\newcommand{\code}[1]{\colorbox{light-gray}{\texttt{#1}}} 

\usepackage{color}

\begin{document}

\title{Empirical Analysis of Software Vulnerabilities Causing Timing Side Channels}

\author{\IEEEauthorblockN{
    M. Mehdi Kholoosi\IEEEauthorrefmark{1}\IEEEauthorrefmark{2}, 
    M. Ali Babar\IEEEauthorrefmark{1}\IEEEauthorrefmark{2},
    Cemal Yilmaz\IEEEauthorrefmark{3}
    }
    \IEEEauthorblockA{\IEEEauthorrefmark{1} School of Computer Science, CREST, The University of Adelaide, Adelaide, Australia} 
    \IEEEauthorblockA{\IEEEauthorrefmark{2} Cyber Security Cooperative Research Centre, Australia}
    \IEEEauthorblockA{\IEEEauthorrefmark{3} Faculty of Engineering and Natural Sciences, Sabanci University, Istanbul, 34956, Turkey}
    Emails: mehdi.kholoosi@adelaide.edu.au, ali.babar@adelaide.edu.au, cemal.yilmaz@sabanciuniv.edu}


\maketitle

\begin{abstract}
Timing attacks are considered one of the most damaging side-channel attacks. These attacks exploit timing fluctuations caused by certain operations to disclose confidential information to an attacker. For instance, in asymmetric encryption, operations such as multiplication and division can cause time-varying execution times that can be ill-treated to obtain an encryption key. Whilst several efforts have been devoted to exploring the various aspects of timing attacks, particularly in cryptography, little attention has been paid to empirically studying the timing attack-related vulnerabilities in non-cryptographic software. By inspecting these software vulnerabilities, this study aims to gain an evidence-based understanding of weaknesses in non-cryptographic software that may help timing attacks succeed. We used qualitative and quantitative research approaches to systematically study the timing attack-related vulnerabilities reported in the National Vulnerability Database (NVD) from March 2003 to December 2022. Our analysis was focused on the modifications made to the code for patching the identified vulnerabilities. We found that a majority of the timing attack-related vulnerabilities were introduced due to not following known secure coding practices. The findings of this study are expected to help the software security community gain evidence-based information about the nature and causes of the vulnerabilities related to timing attacks.
\end{abstract}

\vspace{-2mm}
\section{Introduction}
\label{sec:introduction}

Cyber threats are largely driven by Security Vulnerabilities (SVs) \cite{liu2019deepbalance}. They are characterised as weaknesses in a computer system that attackers utilise to perform malicious actions \cite{dowd2006art}. Recently, a sharp increase has been seen in the number of recorded SVs. This is evident from the Common Vulnerabilities and Exposures (CVE)\cite{CVE}, a database that records publicly released information about security flaws. For instance, the number of registered SVs nearly quadrupled in 2022 compared to 2016 (from 6,454 to 25,226 vulnerabilities). 

The protection of secret data (e.g., passwords, cryptographic keys) in software is a challenging task \cite{cauligi2020constant}. In this regard, cryptographic algorithms and protocols play a critical role in the security of a computer system\cite{lazardoes}. However, any form of SVs in cryptographic implementations (e.g., cryptographic libraries) poses a significant threat to their effectiveness, as it is common for attackers to exploit weaknesses to evade security mechanisms \cite{carre2019cache}.
One of the most viable ways to compromise the security measures of a system is through timing side-channel attacks. By exploiting variations in the time required to execute cryptographic operations (e.g., encryption, decryption), these attacks are able to uncover secrets in a non-invasive way\cite{reparaz2017dude}. Kocher et al.\cite{kocher1996timing} pioneered the notion of timing attacks in 1996. They demonstrated that precise timing measurement of cryptographic operations could reveal the entire private key of various cryptography systems to an attacker. Since then, timing attacks have further evolved and broken numerous major cryptographic implementations \cite{brumley2011remote, bernstein2005cache, ge2018survey, al2013lucky}. Also, there have been many countermeasures proposed against these attacks\cite{ge2018survey}. Timing attacks are particularly concerning because, unlike other side-channel attacks (e.g., electromagnetic attacks\cite{genkin2015get} and power consumption attacks\cite{coron1999resistance}), restricting physical access to the target device is not sufficient to prevent them.  Furthermore, these attacks can be launched remotely\cite{brumley2011remote}, which gives adversaries a wide range of attack options. Finally, timing attacks are known to be nearly untraceable and may only leave suspicious access logs on the targeted computer\cite{jancar2022they}. That is why the extent to which these attacks are launched in the real world is unknown.

Given the devastating consequences of successful timing attacks, several research efforts have been allocated to extensively analyse different aspects of these attacks. However, most of these efforts have focused on studying offense and defense techniques at various levels of cryptography, including algorithms, protocols, implementations, and hardware. We found that no systematic research had been conducted regarding the timing vulnerabilities in non-cryptographic software (i.e., applications that do not implement cryptographic algorithms and protocols). It is equally important to systematically explore and understand the application (i.e., non cryptographic) level vulnerabilities that may contribute to timing attacks since a timing vulnerability is more likely to leak exploitable information the more often it is executed\cite{carre2019cache}. 


To fill this research gap, we performed an empirical study aimed at identifying application-level timing SVs and examining the source code of these vulnerabilities and their fixes to determine the responsible coding mistakes. We selected and analysed vulnerabilities reported over approximately two decades on the National Vulnerability Database (NVD) \cite{NVD}. We performed a thorough inspection of 67 software vulnerabilities that were detected in 56 unique non-cryptographic projects. Our empirical analysis has enabled us to identify and categorise application-level timing vulnerabilities and the coding mistakes that can introduce such vulnerabilities. We assert that this study is the first of its kind, whose findings are expected to provide practitioners and researchers with valuable insights into the nature of timing vulnerabilities and how to reduce the chances of their introduction in software applications by avoiding certain coding mistakes. 


The key contributions of this study are:
\begin{enumerate}
    \item We have carried out a first of its kind empirical study to demonstrate the existence of timing vulnerabilities at the non-cryptographic level.
    \item We have systematically gathered and analysed the relevant data from different sources to identify the secure coding practices that developers can use to avoid the introduction of timing side channels in applications (i.e., non-cryptographic).
    \item We have released a fine-grained dataset of vulnerabilities related to timing attacks for further research\cite{Kholoosi2023}. 
\end{enumerate}
  
\vspace{-1mm}
\section{Background and Related Work}
\label{sec:background}
Our research relates to prior works that have investigated the constant-time coding paradigm. Constant-time programming has become the dominant software-driven countermeasure against these attacks and has been adopted in many major cryptographic implementations. In this programming technique, the behaviour of the code is not dependent on the secret data\cite{ge2018survey}. To fulfill this, it is necessary to ensure that secrets do not influence the control flow (e.g., branch conditions) of the program and its addresses of memory accesses (e.g., array indexes)\cite{jancar2022they}. Moreover, secrets should not affect the inputs of variable-time machine operations, such as integer division\cite{disselkoen2020finding}. In the absence of these protections, an attacker may be able to infer secret data through the analysis of the timing side channel.

The process of writing constant-time code is inherently challenging\cite{reparaz2017dude}. Prior research indicated that even prominent cryptographic implementations that were deemed to be constant-time had timing leakages\cite{pereida2016make}. 
Several tools have been created for automatic constant-time verification\cite{reparaz2017dude}. Disselkoen et al.\cite{disselkoen2020finding} presented Pitchfork, a constant-time verification tool that can analyse cryptographic codes at both primitives and protocol levels.

As opposed to previous studies, which focused on timing attacks and constant-time coding at cryptographic levels (i.e., primitives, protocols, libraries), our study examines timing attack vulnerabilities at the non-cryptography level (i.e., application level). Software at this level is written in high-level programming languages (e.g., Java), and several factors can influence its timing behaviour\cite{ge2018survey}. For example, these languages often provide abstractions and optimisation features to make the code more efficient. However, these features may result in timing discrepancies in the underlying code, thus compromising constant-time execution\cite{barthe2018secure}. Another factor that impacts timing behaviour is human error, which is the primary focus of our study. As far as we know, our paper is the first to identify common coding mistakes made by developers in high-level programming languages that undermine constant-time programming.
\vspace{-2mm}
\section{Study Design}
\label{sec:method}
\begin{figure}[h]
  \centering
  \includegraphics[width=0.95\linewidth]{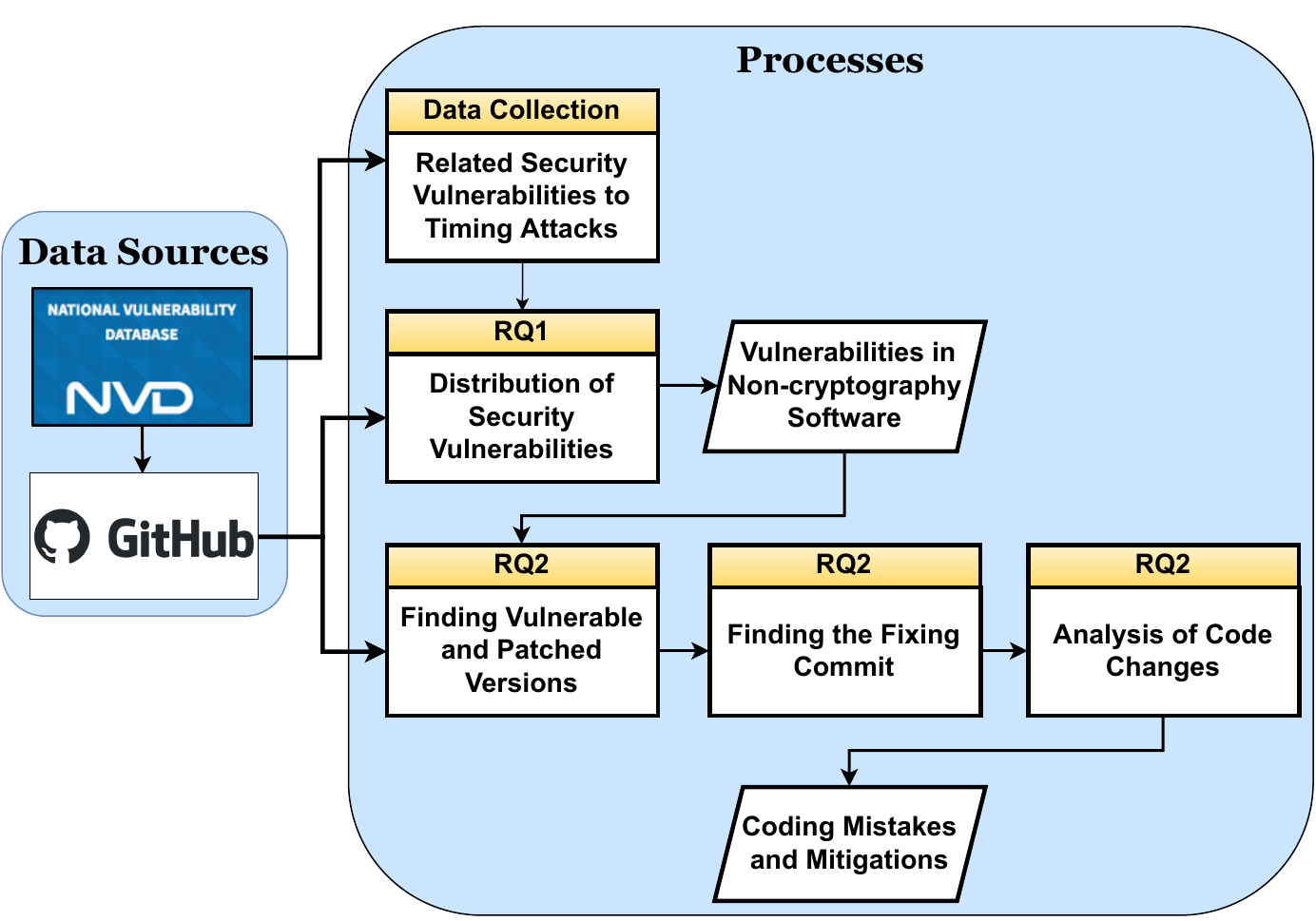}
  \caption{The overall study design.}
  \label{fig:design}
\end{figure}

To understand the characteristics of relevant SVs to timing attacks, we address two Research Questions (RQs). Figure \ref{fig:design} illustrates the overall workflow used to conduct this study. We first introduce our research questions in Section \ref{sec:questions}, then describe the process of collecting and preparing the data used to answer these questions in Section \ref{sec:collection}, and finally present the analysis processes and our findings in Section \ref{sec:analysis}.  
\vspace{-4mm}
\subsection{Research Questions}
\label{sec:questions}
Our investigation is guided by the following RQs: 

\begin{itemize}

    \item \textbf{RQ1: How prevalent are timing attack-related security vulnerabilities in non-cryptographic software?} 
 First, we aim to demonstrate the prevalence of timing attack-related SVs in non-cryptographic software. Additionally, the RQ1 findings would benefit the software security community by shedding light on how these SVs are associated with their affected products (i.e., products that have been considered vulnerable to timing attacks).
    
    \item \textbf{RQ2: (a) What coding mistakes make non-cryptographic software more vulnerable to timing attacks? (b) How do developers patch these software vulnerabilities in real-world projects?} RQ2 findings would give insights to practitioners about the common coding mistakes that increase the success rate of timing attacks. Furthermore, they will unveil what kinds of patches are available to address these vulnerabilities in the source code.
    

    


    
\end{itemize}

\subsection{Data Collection}
\label{sec:collection}

In this study, we used NVD\cite{NVD}, a standards-based vulnerability management data repository. This database was selected specifically because it is considered trustworthy as it is maintained by cybersecurity experts from the National Institute of Standards and Technology (NIST) - a governmental agency of the United States Department of Commerce. It provides a wide range of information regarding known SVs, such as security-related flaws, misconfigurations, affected product names, and impact metrics. Following existing text analysis studies \cite{pranckevivcius2017comparison,rajput2020word}, we developed an iterative data-gathering approach to gain a comprehensive overview of SVs related to timing attacks. 


We started data collection by exact-matching the key phrase ``timing attack" on NVD and then collected the \textit{CVE-ID} (i.e., vulnerability id) and \textit{Description} (i.e., vulnerability summary) fields of 89 SVs. Since other SVs may still exist in NVD that do not contain the phrase ``timing attack" in their \textit{Description} fields, we decided to rerun the search using other related keywords to timing attacks in order to ensure we collected as many related SVs as possible. 
For this reason, we used Natural Language Toolkit (NLTK)\cite{bird2009natural} in the next step to find the most common phrases within the \textit{Description} fields of previously 89 collected SVs. Initially, we performed preprocessing tasks, which included removing stop words (e.g., `a', `the', `is', `and') and punctuations to clean the data. Furthermore, we captured the bigrams and trigrams of all the descriptions and calculated the total appearance frequency of each phrase. The authors of this paper then collectively selected the most relevant key phrases to timing attacks from the list of most popular phrases. Table \ref{tab:Selected keywords} displays the selected key phrases with their frequency.
\vspace{-3mm}
\begin{table}[h]
  \centering
  \caption{Selected key phrases.}
  \label{tab:Selected keywords}
  \begin{tabular}{|p{4cm}|p{1.5cm}|}
    \hline
    \textbf{Key phrase} & \textbf{Frequency}\\
    \hline
    \text{constant time} & 12\\
    \hline
    \text{timing discrepancy} & 10\\
    \hline
    \text{timing difference} & 7\\
    \hline
    \text{HMAC comparison} & 6\\
    \hline
    \text{timing side channel} & 6\\
    \hline
    \text{observable timing discrepancy} & 6\\
    \hline
    \text{constant time comparison} & 4\\
    \hline
    \text{timing issue} & 4\\
    \hline
    \text{timing information} & 3\\
    \hline
    \text{non time constant} & 3\\
    \hline
    \text{timing channel} & 2\\
    \hline
    \text{timing measurement} & 2\\
    \hline
    \text{cache timing} & 2\\
    \hline
    \text{timing leakage} & 2\\
    \hline
    \text{execution time differences} & 2\\
    \hline
\end{tabular}
\end{table}
After finalising the list of keywords, we searched NVD again using each of them and collected the \textit{CVE-ID} and \textit{Description} fields of retrieved SVs. After removing the duplicated SVs (i.e., same CVE-ID), another round of manual analysis was conducted in order to verify that all of the SVs collected were related to timing attacks. To this end, two authors of this paper individually read each vulnerability's \textit{Description} field and assessed whether the selected vulnerability was related to timing attacks. Each vulnerability was labelled as either related or unrelated. 
Following that, the annotators compared their labels, and if disagreements were found, they referred to the \textit{References} section of the vulnerability on NVD. External links are provided in this section to additional resources (e.g., research papers, vendor advisory) that give more information regarding the particular vulnerability of interest. This process was repeated until there was a consensus between the annotators. 
\section{Analysis and Results}
\label{sec:analysis}
\subsection{Distribution of Security Vulnerabilities}
\label{sec:distribution}
 As our study was focused on SVs in non-cryptographic software, we needed to know the distribution of SVs across their \textit{affected product} (i.e., list of products, platforms and/or hardware that are considered to be vulnerable) fields. This is because SVs can exist in a variety of products, including hardware and different kinds of software in a computer system. So, to figure out which SVs happened in non-cryptographic software, we decided to classify all of the collected SVs by the types (e.g., cryptographic library) of their affected products in RQ1. 

\textbf{\textit{RQ1: How prevalent are timing attack-related security vulnerabilities in non-cryptographic software?}}

We categorised the 243 SVs in our dataset based on their \textit{Affected Product} field. If the affected product were unfamiliar, we would refer to its online repository for further information regarding its functionalities. By following this procedure, we assigned each SV from our dataset to the following seven categories of products. We used similar categories to Lazar et al.\cite{lazardoes}.


\begin{itemize}
    \item \textbf{Application.} Refers to non-cryptographic software (e.g., Jenkins).
    \item \textbf{Cryptographic Library.} Refers to the implementation of cryptographic protocols (e.g., OpenSSL).
    \item \textbf{Cryptographic Protocol.} Refers to methods which provide details about how cryptographic algorithms need to be utilised (e.g., TLS).
    \item \textbf{Cryptographic Primitive.} Refers to low-level algorithms in cryptography (e.g., AES).
    \item \textbf{OS.} Refers to the underlying Operating System (e.g., Microsoft Windows).
    \item \textbf{Firmware.} Refers to a single-purposed software that includes machine-level instructions for hardware (e.g., router firmware).
    \item \textbf{Hardware.} Refers to the physical layer of a computer system (e.g., Intel microprocessor).
\end{itemize}


Then, we counted the occurrence of SVs within each category. Table \ref{tab:frequency} displays the frequency distribution of SVs related to timing attacks within each product category.
\vspace{-3mm}
\begin{table}[h]
  \centering
  \caption{Frequency of vulnerabilities in each category of vulnerable products}
  \label{tab:frequency}
  \begin{tabular}{|c|c|c|}
    \hline
    \textbf{Product Category} & \textbf{\# Vulnerabilities} & \textbf{Percentage}\\
    \hline
    Application & 125 & 51.44\\
    Cryptographic Library & 74 & 30.45\\
    Cryptographic Protocol & 1 & 0.41\\
    Cryptographic Primitive & 1 & 0.41\\
    OS & 12 & 4.94\\
    Firmware & 26 & 10.7\\
    Hardware & 4 & 1.65\\
    \hline
    Total & 243 & 100\\
    \hline
\end{tabular}
\end{table}
\vspace{-2mm}
As can be seen, the Application category contains over 51\% of SVs. Developers in this category typically do not have expertise in cryptography
\cite{lazardoes}. Further details about these SVs are provided later in Section \ref{sec:secure-coding}.

We observed that 30\% of collected SVs belong to the Cryptographic Library category. Cryptographic libraries are popular with software developers since they provide varying degrees of security that can be integrated into applications for data transmission and storage. They are written by expert developers with profound cryptography knowledge \cite{jancar2022they}, and any errors in their implementation adversely affect many applications that rely on these libraries. SVs in this category have been studied extensively by cryptography researchers\cite{al2013lucky,brumley2011remote}.

In the Cryptographic Protocol and Cryptographic Primitive categories, we have collected one timing attack-related vulnerability per category. The vulnerability in the Cryptographic Protocol category (CVE-2013-0169) is related to the Lucky Thirteen attack \cite{al2013lucky}. It is a cryptographic timing attack based on a detailed timing analysis of decryption processing in the Transport Layer Security (TLS) protocol. This vulnerability is an example of a dependency vulnerability that disseminates to other categories through dependent software products. The discovery of these vulnerabilities may threaten the security of dependent software. For instance, AlFardan et al.\cite{al2013lucky} demonstrated in their experiments that TLS implementations in cryptographic libraries such as GnuTLS, Network Security Services (NSS), CyaSSL, and BouncyCastle are vulnerable to the Lucky Thirteen attack. The aforementioned software products indeed belong to the Cryptographic Library category, which is a higher level category than the Cryptographic Protocol category. In addition, they reported that this vulnerability affected the Opera browser (CVE-2013-1618) - a software product belonging to the Application category.

The vulnerability in the Cryptographic Primitive category (CVE-2005-1797) is related to cache-timing attacks. These attacks represent a specific category of side-channel attacks that exploit the cache behaviour of contemporary computing systems to acquire knowledge about encryption keys. Bernstein \cite{bernstein2005cache} investigated the vulnerability of AES encryption to cache-timing attacks and demonstrated that encryption operations conducted with an AES key result in particular cache access patterns, which can be exploited to deduce information about the key. 

There were 12 (4.94\%) and 26 (10.7\%) SVs in the OS and Firmware categories, respectively. However, we could not gather more information about these vulnerabilities because they were closed-source products or the information provided regarding the vulnerabilities was quite limited. 

  Among the SVs in our dataset, four belong to the Hardware category. The Meltdown \cite{Lipp2018meltdown} (CVE-2017-5754) and Spectre \cite{Kocher2018spectre} (CVE-2017-5715, CVE-2017-5753) attacks exploited transient execution CPU vulnerabilities and impacted a wide range of modern processors from Intel, AMD, and the ARM family. These SVs have been attributed to design decisions made by hardware manufacturers during the implementation of the speculative execution and branch prediction mechanisms\cite{Lipp2018meltdown,Kocher2018spectre}. The fourth vulnerability in this category is PortSmash \cite{aldaya2019port} (CVE-2018-5407) which impacts processors that run on an SMT (Simultaneous Multithreading) architecture where multiple threads can be executed simultaneously on a single CPU core. As a proof of concept, Aldaya et al.\cite{aldaya2019port} exploited Intel Hyper-Threading technology and performed their timing side-channel attack on Intel Skylake and Kaby Lake architectures. 

Since our study aims to analyse the relevant coding mistakes in non-cryptographic software, in Section \ref{sec:secure-coding}, we solely focus on the Application category, which indeed turns out to have the most number of timing attack-related SVs. 
\vspace{-1mm}
\subsection{Source-Code level Analysis}
\label{sec:secure-coding}
Source code level analysis involves systematically reviewing the source code of a software application in order to gain a deeper understanding of its structure, functions, and limitations. In this section, we sought to identify relevant code changes that patched the vulnerability and then utilise them to identify common coding mistakes and mitigation techniques.

\textit{\textbf{RQ2: (a) What coding mistakes make non-cryptographic software more vulnerable to timing attacks? (b) How do developers patch these software vulnerabilities in real-world projects?}}

To perform this analysis, we had to have access to the source code of the products that were considered vulnerable (i.e., \textit{Affected Product}). For this purpose, we began locating the corresponding repositories on GitHub or other software repositories for each of the vulnerabilities within the Application category of our dataset. We excluded vulnerabilities from our analysis whose affected products were closed-source (25 entries). As a result, we were left with 100 vulnerabilities in the Application category. The following steps were taken for each of the remaining vulnerabilities:

\subsubsection {\textbf{Finding the vulnerable and patched versions of the affected product}}\label{Sec:Vuln-and-Fixed} We first looked into the \textit{Affected Product} section of NVD for each vulnerability. Bao et al. \cite{bao2022v} reported some inaccuracy regarding the information in this section of NVD. As an example, a certain version can be marked as vulnerable while it is not actually vulnerable. The findings of this studies led us to consider multiple sources of information to increase the reliability of the data we extracted. Since the product's advisory page is the official source of information, we mainly relied on it to identify the patched and vulnerable versions.

\subsubsection {\textbf{Finding the fixing commit of the affected product}} To find the exact code changes that patched the vulnerability, we first needed to determine which commit was the fixing commit(s). The reason for this is that most large OSS (Open Source Software) projects contain multiple commits within a single version. A link to the fixing commit(s) is usually included in the \textit{References} section of each vulnerability in NVD. If a direct link was not found, we manually looked for the fixing commit(s) in the corresponding repository we previously located. As part of this process, we searched the repository with \textit{CVE-ID} as a keyword because repository maintainers usually use it in various sections (e.g., conversations, commit descriptions, code comments) for future reference. If \textit{CVE-ID} failed to provide any results, we searched the repository using keywords from Table\ref{tab:Selected keywords} to identify fixing commit(s). 
This approach enabled us to locate fixing commits for 67 out of 100 vulnerabilities. Due to the insufficient information provided, we could not pinpoint the relevant commits and code changes for the remaining 33 vulnerabilities. We marked these vulnerabilities with ``insufficient info" and excluded them from the rest of the analysis. 
    

\subsubsection {\textbf{Analysis of code changes}} \label{Sec:saving-information} Following the identification of the fixing commit in the previous step, we examined all the information available for each vulnerability. We used a similar approach to Croft et al.\cite{croft2023data} to manually analyse code changes related to a vulnerability. Initially, we focused on the sections \textit{Description} and \textit{References} of NVD. The links in the \textit{References} section typically point to valuable information from release notes and official advisory pages of the software product. Furthermore, we concentrated on the information available at the source-code level. Fixing commit description and code comments were taken into consideration, as developers often use these to describe the functionality of code and the context in which it was changed. Upon acquiring a comprehensive understanding of the vulnerability, we examined the changed lines, along with the entire fixing commit code, to determine the underlying cause of the change. We recorded our observations as coding mistakes and the nature of the change for each vulnerability. We followed this approach to gradually build a taxonomy of common mistakes and mitigations.

This 3-step manual analysis consumed over 120 hours of effort and was carried out by the first author, who had three years of experience in software security. In the course of this process, several weekly meetings were held with the other two authors to reduce bias and inaccuracies. In these meetings, we randomly performed the same manual process for several vulnerabilities in different programming languages.


\subsection{Results}
We encountered two categories of coding mistakes during our manual analysis. Table \ref{tab:frequency-mistakes} displays the frequency of vulnerabilities in each category of coding mistakes.
\vspace{-3mm}
\begin{table}[h]
  \centering
  \caption{Frequency of coding mistakes in each category}
  \label{tab:frequency-mistakes}
  \begin{tabular}{|c|c|c|}
    \hline
    \textbf{Category of Coding Mistake} & \textbf{\# Vulnerabilities} & \textbf{Percentage}\\
    \hline
    Unsafe Comparison & 60 & 89.55\\
    User Enumeration Issues & 7 & 10.45\\
    \hline
    Total & 67 & 100\\
    \hline
\end{tabular}
\end{table}

\subsubsection {\textbf{Unsafe Comparison}} The most common coding mistake that we found was lack of constant-time comparison, which is a security technique that aims to ensure the consistency and independence of the processing time associated with comparison operations. Its origins are in computer security, where the processing time of a comparison operation between two values can lead to the disclosure of sensitive information through side-channel attacks, such as timing attacks \cite{reparaz2017dude}. A fixed number of operations for each comparison is guaranteed in a constant-time comparison, and regardless of the input values, an attacker cannot deduce any information about the secret from execution time \cite{bernstein2019fast}. Based on our source code analysis, we observed that 90\% (60 out of 67) of all vulnerabilities in the Application category are associated with unsafe comparison operations. We describe our observations for Java, PHP, and C programming languages in the following. Additionally, to illustrate the popularity level of each product, we display the number of \textit{Stars} and \textit{Forks} within their respective GitHub repositories. We provide further details about popularity level in Section \ref{sec:threats}.
   
Table \ref{tab:java} presents a selection of vulnerabilities in Java-based applications that are susceptible to timing attacks due to unsafe comparison operations.
\vspace{-4mm}
\begin{table}[h]
\centering
\caption{A selection of Related software vulnerabilities with not constant-time comparison in Java.}
  \label{tab:java}
\begin{tabular}{|l|l|l|l|}
\hline
\rowcolor[HTML]{FFFFC7} 
\multicolumn{1}{|c|}{\cellcolor[HTML]{FFFFC7}\textbf{CVE-ID}} & \multicolumn{1}{c|}{\cellcolor[HTML]{FFFFC7}\textbf{Product}} & \multicolumn{1}{c|}{\cellcolor[HTML]{FFFFC7}\textbf{Star}} & \multicolumn{1}{c|}{\cellcolor[HTML]{FFFFC7}\textbf{Fork}} \\ \hline
CVE-2020-1926                                                 & hive                                                          & 34.9k                                                      & 26.6k                                                      \\
CVE-2021-38153                                                & kafka                                                         & 24.1k                                                      & 12.3k                                                      \\
CVE-2020-2102                                                 & jenkins                                                       & 20.2k                                                      & 7.9k                                                       \\
CVE-2020-2101                                                 & jenkins                                                       & 20.2k                                                      & 7.9k                                                       \\
\hline
\end{tabular}
\end{table}
We tracked the code changes that patched the vulnerability in these products and observed that \code{Array.equals()} and \code{String.equals()} methods are not considered timing attack safe. When the code reaches a point where it needs to compare a secret, \code{MessageDigest.isEqual()} is the preferred method in the patches. Listing \ref{lst:listing-java1} displays a code snippet (related to CVE-2021-38153) from one of the components in Kafka (i.e., distributed data streaming platform) in which two keys need to be compared in order to authenticate the user. The vulnerability has been patched by using the \code{MessageDigest.isEqual()} (line 8) method instead of \code{Array.equals()}. The added and deleted lines are highlighted in green and red, respectively. 
    
\begin{lstlisting}[language=Java, label={lst:listing-java1}, caption=Example of unsafe comparison in Java\, adapted from Kafka repository \cite{java-github}., escapeinside=@@]
  import java.security.InvalidKeyException;
@\colorbox{lightgreen}{+}@import java.security.MessageDigest;
...
      byte[] expectedStoredKey = scramCredential.storedKey();
      byte[] clientSignature = formatter.clientSignature(expectedStoredKey, clientFirstMessage, serverFirstMessage, clientFinalMessage);
      byte[] computedStoredKey = formatter.storedKey(clientSignature, clientFinalMessage.proof());
@\colorbox{lightcoral}{-}@    if (! @\colorbox{lightcoral}{Arrays.equals}@(computedStoredKey, expectedStoredKey))
@\colorbox{lightgreen}{+}@    if (!@\colorbox{lightgreen}{MessageDigest.isEqual}@(computedStoredKey, expectedStoredKey))
          throw new SaslException("Invalid client credentials");
   } catch (InvalidKeyException e) {
       throw new SaslException("Sasl client verification failed", e);
...
\end{lstlisting}
In the Java Cryptography Architecture (JCA), the \code{MessageDigest} class generates secure hashes of data, also known as message digests. In this class, message digest algorithms, such as MD5 and SHA-256, are provided for applications. The \code{MessageDigest.isEqual()} method is a utility method used for comparing the results of two \code{MessageDigest} objects. It returns a boolean value of true if the hash values generated by the two \code{MessageDigest} objects are equal, and false otherwise. This means that unlike \code{Array.equals()} and \code{String.equals()} methods, \code{MessageDigest.isEqual()} will not return immediately and continue the comparison operation byte by byte to the end irrespective of the first byte being different. This behaviour makes this method secure but less efficient.

 In Table \ref{tab:php}, we display a selection of vulnerabilities in PHP-based applications that are susceptible to timing attacks.
\vspace{-4mm}
\begin{table}[h]
\centering
\caption{A selection of Related software vulnerabilities with not constant-time comparison in PHP.}
  \label{tab:php}
\begin{tabular}{|l|l|l|l|}
\hline
\rowcolor[HTML]{FFFFC7} 
\multicolumn{1}{|c|}{\cellcolor[HTML]{FFFFC7}\textbf{CVE-ID}} & \multicolumn{1}{c|}{\cellcolor[HTML]{FFFFC7}\textbf{Product}} & \multicolumn{1}{c|}{\cellcolor[HTML]{FFFFC7}\textbf{Star}} & \multicolumn{1}{c|}{\cellcolor[HTML]{FFFFC7}\textbf{Fork}} \\ \hline
CVE-2017-14775                                                & laravel                                                       & 28.8k                                                      & 9.9k                                                       \\
CVE-2019-18887                                                & symfony                                                       & 28k                                                        & 9k                                                         \\
CVE-2015-5730                                                 & wordpress                                                     & 17.1k                                                      & 12k                                                        \\
CVE-2016-2041                                                 & phpMyAdmin                                                    & 6.4k                                                       & 3.3k                                                       \\
\hline
\end{tabular}
\end{table}
 Based on the code changes that patched these vulnerabilities, we found that native comparison operators in PHP (\code{==, !=, ===, !==}) are regarded as unsafe against timing attacks when used to compare sensitive data such as cryptographic keys and passwords. In the provided patches, we observed that developers used the \code{hash\_equals()} function (in PHP versions greater than 5.6.0) to patch these vulnerabilities. This built-in function is suitable for comparing two strings in a time-constant manner. It accepts two input string arguments and evaluates their lengths in the first instance. If the strings differ in length, the function immediately returns false. Otherwise, the function conducts a constant-time comparison of the two values through a byte-by-byte comparison method and returns a boolean value of true if the values are identical or false if they are not. 

% ...


% ...
Listing \ref{lst:php2-unsafe} displays a code snippet (related to CVE-2015-5730) from WordPress (i.e., web development platform) repository. It was indicated in release notes of WordPress that an attacker could have leveraged this vulnerability by a timing attack to lock a post and prevent it from being edited \cite{wordpress.org-documentation}.

\begin{lstlisting}[language=PHP, label={lst:php2-unsafe}, caption=Example of unsafe comparison in PHP\, adapted from WordPress repository \cite{php2-wordpress}., escapeinside=@@]
...
@\colorbox{lightcoral}{-}@      if ( $this->get_instance_hash_key( $decoded ) @\colorbox{lightcoral}{!==}@ $value['instance_hash_key'] ) {
@\colorbox{lightgreen}{+}@      if ( ! @\colorbox{lightgreen}{hash\_equals}@( $this->get_instance_hash_key( $decoded ), $value['instance_hash_key'] ) ) {
            return null;
        }
...
\end{lstlisting}

Table \ref{tab:c} lists vulnerabilities in applications written in C language that are susceptible to timing attacks as a result of unsafe comparison operations.
\begin{table}[h]
\centering
\caption{Related software vulnerabilities with not constant-time comparison in C.}
  \label{tab:c}
\begin{tabular}{|l|l|l|l|}
\hline
\rowcolor[HTML]{FFFFC7} 
\multicolumn{1}{|c|}{\cellcolor[HTML]{FFFFC7}\textbf{CVE-ID}} & \multicolumn{1}{c|}{\cellcolor[HTML]{FFFFC7}\textbf{Product}} & \multicolumn{1}{c|}{\cellcolor[HTML]{FFFFC7}\textbf{Star}} & \multicolumn{1}{c|}{\cellcolor[HTML]{FFFFC7}\textbf{Fork}} \\ \hline
CVE-2013-2061                                                 & openvpn                                                       & 8k                                                         & 2.6k                                                       \\
CVE-2017-2624                                                 & xorg-server                                                   & 146                                                        & 56                                                         \\
CVE-2020-11683                                                & at91bootstrap                                                 & 98                                                         & 221                                                        \\
CVE-2021-37848                                                & barebox                                                       & 81                                                         & 40                                                         \\
CVE-2021-37847                                                & barebox                                                       & 81                                                         & 40                                                         \\ \hline
\end{tabular}
\end{table}
As per the patches provided for these vulnerabilities, \code{memcmp()} and \code{strncmp()} are not secure against timing attacks since their comparisons stop once the first difference is encountered. Contrary to other languages, C and C++ developers implement their own functions to ensure safe comparisons. For instance, Listing \ref{lst:C1-unsafe} displays a code snippet (related to CVE-2013-2061) from OpenVPN (i.e., network software) repository. The openvpn\_decrypt function in crypto.c of OpenVPN 2.3.0 and earlier uses the \code{memcmp()} function (line 3) for comparing HMAC (Hash-based message authentication code) tokens. The non-constant-time nature of this function may enable remote attackers to obtain sensitive information through a timing attack.

\begin{lstlisting}[language=C, label={lst:C1-unsafe}, caption=Example of unsafe comparison in C\, adapted from OpenVPN repository \cite{C1-github}., escapeinside=@@]
...
	  /* Compare locally computed HMAC with packet HMAC */
@\colorbox{lightcoral}{-}@  if (@\colorbox{lightcoral}{memcmp}@ (local_hmac, BPTR (buf), hmac_len))
@\colorbox{lightgreen}{+}@  if (@\colorbox{lightgreen}{memcmp\_constant\_time}@ (local_hmac, BPTR (buf), hmac_len))
	    CRYPT_ERROR ("packet HMAC authentication failed");

	  ASSERT (buf_advance (buf, hmac_len));
...
\end{lstlisting}
To mitigate this vulnerability, a new function has been implemented by developers and included in the patch to perform the comparison in constant-time. The implementation of \code{memcmp\_constant\_time()} can be seen in Listing \ref{lst:C2-unsafe}.

\begin{lstlisting}[language=C, label={lst:C2-unsafe}, caption=Implementation of a constant-time comparison function in C adapted from OpenVPN repository \cite{C1-github}., escapeinside=@@]
static int
memcmp_constant_time (const void *a, const void *b, size_t size) {
  const uint8_t * a1 = a;
  const uint8_t * b1 = b;
  int ret = 0;
  size_t i;

  for (i = 0; i < size; i++) {
      ret |= *a1++ ^ *b1++;
  }

  return ret;
}
\end{lstlisting}
 This function is designed to compare two blocks of memory (\code{a} and \code{b}) in constant-time. To achieve this, in line 9, the function uses the bitwise XOR operator (\code{\^}) to compare each byte of the memory blocks. The XOR operation compares each bit of two input values and returns \code{1} in each bit position where the corresponding bits of the two operands are different and \code{0} where they are the same. The resulting bits from each byte comparison are then ORed together to form a final result (\code{ret}), which indicates whether the two memory blocks are equal or not. This approach ensures that the execution time of the function is constant and independent of the values being compared, which prevents timing side channels from revealing information about the memory blocks. We discovered similar implementations of constant-time comparison functions in the patches provided for the remaining vulnerabilities listed in Table \ref{tab:c}. 
Additionally, we observed that the same technique (i.e., bitwise XOR operation) was used in implementing constant-time comparison functions in C++ \cite{cpp2-github}.

Table \ref{tab:11langs} summarises our observations regarding safe and unsafe comparisons in 11 different programming languages.
We have provided the complete list of vulnerabilities and their fixing commit links in our released dataset\cite{Kholoosi2023} of vulnerabilities related to timing attacks.


\begin{table*}[]
\centering
\caption{Safe and unsafe comparisons in different programming languages.}
\label{tab:11langs}
\begin{tabular}{|c|l|l|l|}
\hline
\rowcolor[HTML]{FFFFC7} 
\textbf{Language}                                    & \multicolumn{1}{c|}{\cellcolor[HTML]{FFFFC7}\textbf{Unsafe comparison}}                                               & \multicolumn{1}{c|}{\cellcolor[HTML]{FFFFC7}\textbf{Constant-time comparison function}} & \multicolumn{1}{c|}{\cellcolor[HTML]{FFFFC7}\textbf{Available in}}\\ \hline
\rowcolor[HTML]{DAE8FC} 
Java                                                 & \begin{tabular}[c]{@{}l@{}}Arrays.equals(),\\ String.equals()\end{tabular}                                             & MessageDigest.isEqual()                                                                  & built-in class in the java.security package                                                  \\ \hline
PHP                                                  & ==, !=, ===, !==                                                                                                       & hash\_equals()                                                                           & built-in function in PHP                                           \\ \hline
\rowcolor[HTML]{DAE8FC} 
\cellcolor[HTML]{DAE8FC}                             & \cellcolor[HTML]{DAE8FC}                                                                                               & constant\_time\_compare()                                                                & utils.crypto module in Django framework                             \\ \cline{3-4} 
\rowcolor[HTML]{DAE8FC} 
\multirow{-2}{*}{\cellcolor[HTML]{DAE8FC}Python}     & \multirow{-2}{*}{\cellcolor[HTML]{DAE8FC}==, !=}                                                                       & compare\_digest()                                                                        & hmac module (built-in)                                             \\ \hline
C                                                    & \begin{tabular}[c]{@{}l@{}}memcmp(),\\ strncmp()\end{tabular}                                                          & Self-implemented                                                                                         & NA                                                  \\ \hline
\rowcolor[HTML]{DAE8FC} 
\cellcolor[HTML]{DAE8FC}                             & \cellcolor[HTML]{DAE8FC}                                                                                               & Rack::Utils.secure\_compare()                                                            & Rack (low-level interface)                                         \\ \cline{3-4} 
\rowcolor[HTML]{DAE8FC} 
\multirow{-2}{*}{\cellcolor[HTML]{DAE8FC}Ruby}       & \multirow{-2}{*}{\cellcolor[HTML]{DAE8FC}==, !=}                                                                       & ActiveSupport::SecurityUtils.secure\_compare()                                           & Rail (web framework)                                               \\ \hline
C++                                                  & ==, !=                                                                                                                 & Self-implemented                                                                                        & NA                                                                                                               \\ \hline
\rowcolor[HTML]{DAE8FC} 
\cellcolor[HTML]{DAE8FC}                             & \cellcolor[HTML]{DAE8FC}                                                                                               & crypto.timingSafeEqual()                                                                 & crypto module in Node.js                                                                                   \\ \cline{3-4} 
\rowcolor[HTML]{DAE8FC} 
\cellcolor[HTML]{DAE8FC}                             & \cellcolor[HTML]{DAE8FC}                                                                                               & crypto.createHash()                                                                      & crypto module in Node.js                                                                                  \\ \cline{3-4} 
\rowcolor[HTML]{DAE8FC} 
\multirow{-3}{*}{\cellcolor[HTML]{DAE8FC}JavaScript} & \multirow{-3}{*}{\cellcolor[HTML]{DAE8FC}\begin{tabular}[c]{@{}l@{}}  ===, !==, ==, !=\end{tabular}} & scmp()                                                                                   & scmp package (third-party library)                                                                                                \\ \hline
Go                                                   & ==, !=                                                                                                                 & ConstantTimeCompare()                                                                    & crypto package (built-in)                                                                                                       \\ \hline
\rowcolor[HTML]{DAE8FC} 
Perl                                                 &  eq, ne                                                                                                         & Crypt::Util::constant\_time\_eq()                                                        & Crypt::Util (third-party module)                                                                                                  \\ \hline
Lua                                                  & ==, $\sim$=                                                                                                            & secure\_equals()                                                                         & util.hashes (third-party module)                                                                                                \\ \hline
\rowcolor[HTML]{DAE8FC} 
Rust                                                 & ==, !=                                                                                                                 & constant\_time\_eq()                                                                     & constant\_time\_eq (third-party crate)                                                                                             \\ \hline
\end{tabular}
\end{table*}




\subsubsection {\textbf{User Enumeration Issues}}
User enumeration attack is a type of enumeration attack that targets login pages in an attempt to identify valid usernames. This is accomplished by sending login requests with a list of potential usernames and analysing the error messages returned by the system. In certain circumstances, combining user enumeration with timing attacks may lead to more successful attacks. An attacker can use timing attacks to exploit differences in system response times to infer whether a username is valid. The information thus acquired can then be used to launch additional attacks.

Table \ref{tab:enumeration-7} displays vulnerabilities in products that are susceptible to a mix of timing and user enumeration attacks. These seven vulnerabilities originated from observable timing discrepancies on the login forms. According to the patches provided for these vulnerabilities, the applications took considerably longer to respond to login attempts with a valid username and an invalid password than with an invalid username and password. Thus, an attacker could leverage timing attacks to find valid usernames by attempting to log in and assessing the time it takes to evaluate each login attempt for valid and invalid usernames.
\begin{table}[h]
\centering
\caption{Related software vulnerabilities with user enumeration issues.}
\label{tab:enumeration-7}
\begin{tabular}{|l|l|l|l|l|}
\hline
\rowcolor[HTML]{FFFFC7} 
\multicolumn{1}{|c|}{\cellcolor[HTML]{FFFFC7}\textbf{CVE-ID}} & \multicolumn{1}{c|}{\cellcolor[HTML]{FFFFC7}\textbf{Product}} & \multicolumn{1}{c|}{\cellcolor[HTML]{FFFFC7}\textbf{Star}} & \multicolumn{1}{c|}{\cellcolor[HTML]{FFFFC7}\textbf{Fork}} & \multicolumn{1}{c|}{\cellcolor[HTML]{FFFFC7}\textbf{Language}} \\ \hline
CVE-2022-34174                                                & jenkins                                                       & 20.2k                                                      & 7.9k                                                       & Java                                                           \\
CVE-2016-0762                                                 & tomcat                                                        & 6.5k                                                       & 4.4k                                                       & Java                                                           \\ \hline
CVE-2016-2513                                                 & django                                                        & 69.1k                                                      & 28.7k                                                      & Python                                                         \\
CVE-2022-34623                                                & mealie                                                        & 3k                                                         & 331                                                        & Python                                                         \\
CVE-2017-8342                                                 & radicale                                                      & 2.7k                                                       & 386                                                        & Python                                                         \\ \hline
CVE-2020-11063                                                & typo3                                                         & 909                                                        & 602                                                        & PHP                                                            \\ \hline
CVE-2021-38562                                                & request\_tracker                                              & 716                                                        & 215                                                        & Perl                                                           \\ \hline
\end{tabular}
\end{table}

We observed that one of the common mitigation strategies is to avoid stopping the verification process abruptly. Even if the provided username is invalid, developers verified the invalid password with a synthetic password in order to prevent timing discrepancies \cite{enum1-github}.
For instance, Listing \ref{lst:java-enumerate} displays a code snippet (patch for CVE-2022-34174) from Jenkins repository. The code snippet is a Java method that implements user authentication by checking the validity of the provided username and password against an existing record in a data store (line 4). After applying the patch, when a provided username is invalid (line 9), the method uses a \code{MultiPasswordEncoder} object named \code{PASSWORD\_ENCODER}(line 11) to encode and verify passwords, and employs a precomputed encoded value called \code{ENCODED\_INVALID\_USER\_PASSWORD} (line 24) to intentionally waste time, in order to prevent timing attacks from distinguishing between existing and non-existing users. If the provided username exists, the method checks if the associated password is correct (line 14); if not, it throws a \code{BadCredentialsException}(line 15). The \code{generatePassword()} method (line 25) generates a random password of length 20, which is used to initialise the value of \code{ENCODED\_INVALID\_USER\_PASSWORD} in line 24. 

\begin{lstlisting}[language=Java, label={lst:java-enumerate}, caption=Fixing commit for CVE-2022-34174\, adapted from Jenkins repository \cite{enum1-github}., escapeinside=@@]
...
@\colorbox{lightgreen}{+}@ import java.util.Random;
...
      protected UserDetails authenticate2(String username, String password) throws AuthenticationException {
@\colorbox{lightcoral}{-}@       Details u = load(username);
@\colorbox{lightgreen}{+}@       Details u;
@\colorbox{lightgreen}{+}@       try {
@\colorbox{lightgreen}{+}@           u = load(username);
@\colorbox{lightgreen}{+}@       } catch (UsernameNotFoundException ex) {
@\colorbox{lightgreen}{+}@           // Waste time to prevent timing attacks distinguishing existing and non-existing user
@\colorbox{lightgreen}{+}@           PASSWORD_ENCODER.matches(password, ENCODED_INVALID_USER_PASSWORD);
@\colorbox{lightgreen}{+}@           throw ex;
@\colorbox{lightgreen}{+}@       }
          if (!u.isPasswordCorrect(password)) {
              throw new BadCredentialsException("Bad credentials");
          }
...
      public static final MultiPasswordEncoder PASSWORD_ENCODER = new MultiPasswordEncoder();

@\colorbox{lightgreen}{+}@    /**
@\colorbox{lightgreen}{+}@     * This value is used to prevent timing discrepancies when trying to authenticate with an invalid username
@\colorbox{lightgreen}{+}@     * compared to just a wrong password. If the user doesn't exist, compare the provided password with this value.
@\colorbox{lightgreen}{+}@     */
@\colorbox{lightgreen}{+}@    private static final String ENCODED_INVALID_USER_PASSWORD = PASSWORD_ENCODER.encode(generatePassword());
@\colorbox{lightgreen}{+}@    private static String generatePassword() {
@\colorbox{lightgreen}{+}@        String password = new Random().ints(20, 33, 127).mapToObj(i -> (char) i)
@\colorbox{lightgreen}{+}@                .collect(StringBuilder::new, StringBuilder::appendCodePoint, StringBuilder::append).toString();
@\colorbox{lightgreen}{+}@        return password;
@\colorbox{lightgreen}{+}@    }
...
\end{lstlisting}
Another observed technique for preventing timing oracles and user enumeration is the addition of a random timer during the authentication process \cite{enum3-github}.
\vspace{-2mm}
\section{Discussion}
\label{sec:discussion}
This study aimed at empirically investigating non-cryptographic software weaknesses that may assist timing attacks in succeeding. Through RQ1, we first gained a broad view of timing attack-related SVs across various categories of vulnerable products. Furthermore, through RQ2, we evidentially identified the common coding mistakes that application (i.e., non-cryptographic) developers make regarding these attacks. It is important to clarify that in this study, we were not implying that timing vulnerabilities in non-cryptographic software are more significant than those in cryptographic software. As we mentioned in previous sections, due to the complex nature of timing attacks, it is impossible to guarantee that a piece of code is not vulnerable to these attacks. To this end, we wanted to emphasise that timing leakages should be minimised at all levels of coding, including non-cryptography.
We present the following observations from our empirical analysis of software vulnerabilities related to timing attacks:
\begin{itemize}
    \item{Application developers made repetitive coding mistakes (i.e., unsafe comparison operations), which indicates a lack of awareness of timing attacks.}
    \item{Whenever the code is required to deal with sensitive information (e.g., user authentication), time-constant comparison operations should be used to perform the verification.}
    \item{The verification process on login forms should not be skipped, even in the event of a major error (e.g., incorrect username). The identified mitigation techniques are verifying a synthetic password and adding a random timer.}
\end{itemize}
\vspace{-2mm}
\subsection{Implications for Developers and Researchers} We provide application developers with insight into the scope of timing attacks. Moreover, we suggest secure coding practices that can enhance the security level of their code against the growing threat of such attacks. Specifically, we identified constant-time comparison functions in 11 programming languages, which can be leveraged to improve code resistance to these attacks. Additionally, we outlined strategies for addressing user enumeration issues, which are a known attack vector for timing attacks.
For researchers, we have identified some promising research directions. The results of this study may be used as a basis for a user study with application developers to assess their level of awareness about the secure coding practices. An alternative direction would be to utilise the software vulnerability dataset generated in this research to develop a framework for detecting timing attack vulnerabilities at the non-cryptography level to assist application developers.
\vspace{-3mm}
\section{Threats to Validity}
\label{sec:threats}
\textit{Construct Validity:} Our manual source code analysis may be biased or inaccurate. However, following the previously published studies \cite{croft2023data, croft2022noisy} has helped us minimise such threats and confirm the validity of our methodology.

\textit{Internal Validity:} The outcomes of our investigation may be influenced by the correctness of the analysed patches. Developer expertise is crucial to providing patches that address the intended vulnerability without introducing new threats. Gousios et al.\cite{gousios2015work} demonstrated that popular GitHub repositories have a rigorous peer review process for accepting code contributions. The majority of patches analysed in this study are from popular GitHub repositories whose maintainers are strongly familiar with the coding style of the project. To lessen the aforementioned threat, we reported the Stars and Forks counts of analysed repositories in the vulnerability tables throughout the paper.

\textit{External Validity:} 
Another possible concern is that the number of software vulnerabilities in each programming language may affect the completeness of derived conclusions
. For instance, we examined only one software vulnerability in less popular programming languages such as Go, Perl, Rust, and Lua. More vulnerability information sources are required to draw a generic conclusion. 
We are also aware that there may be other relevant software vulnerabilities that could not be included in our dataset for various reasons, such as insufficient disclosed information or that the software was not open-source.
However, we have systematically searched for related vulnerabilities in the NVD dataset, which is widely used and respected among cybersecurity professionals and contains more than 218k reports about known SVs in software and hardware. 
\vspace{-4mm}
\section{Conclusion}
\label{sec:conclusion}
In this paper, we have conducted an empirical study of 243 security vulnerabilities related to timing attacks. We investigated the prevalence of these vulnerabilities in various categories of affected products. In particular, to improve the security of non-cryptographic software against timing attacks, we performed a thorough source code analysis of 67 software vulnerabilities in 11 programming languages. Through this study, we have revealed that developers who are not cryptography experts tend to make two specific types of coding mistakes (i.e., unsafe comparisons and user enumeration issues) that make their code more vulnerable to these attacks. We found that adherence to a taxonomy of known secure coding practices could have prevented the majority of software vulnerabilities related to timing attacks.
\vspace{-2mm}
\section*{Acknowledgment}
This work has been supported by the Cyber Security Cooperative Research Centre Limited whose activities are partially funded by the Australian Government’s Cooperative Research Centre Programme.
\bibliographystyle{IEEEtran}
\bibliography{main}

\begin{thebibliography}{10}
\providecommand{\url}[1]{#1}
\csname url@samestyle\endcsname
\providecommand{\newblock}{\relax}
\providecommand{\bibinfo}[2]{#2}
\providecommand{\BIBentrySTDinterwordspacing}{\spaceskip=0pt\relax}
\providecommand{\BIBentryALTinterwordstretchfactor}{4}
\providecommand{\BIBentryALTinterwordspacing}{\spaceskip=\fontdimen2\font plus
\BIBentryALTinterwordstretchfactor\fontdimen3\font minus
  \fontdimen4\font\relax}
\providecommand{\BIBforeignlanguage}[2]{{%
\expandafter\ifx\csname l@#1\endcsname\relax
\typeout{** WARNING: IEEEtran.bst: No hyphenation pattern has been}%
\typeout{** loaded for the language `#1'. Using the pattern for}%
\typeout{** the default language instead.}%
\else
\language=\csname l@#1\endcsname
\fi
#2}}
\providecommand{\BIBdecl}{\relax}
\BIBdecl

\bibitem{liu2019deepbalance}
S.~Liu, G.~Lin, Q.-L. Han, S.~Wen, J.~Zhang, and Y.~Xiang, ``Deepbalance:
  Deep-learning and fuzzy oversampling for vulnerability detection,''
  \emph{IEEE Transactions on Fuzzy Systems}, vol.~28, no.~7, pp. 1329--1343,
  2019.

\bibitem{dowd2006art}
M.~Dowd, J.~McDonald, and J.~Schuh, \emph{The art of software security
  assessment: Identifying and preventing software vulnerabilities}.\hskip 1em
  plus 0.5em minus 0.4em\relax Pearson Education, 2006.

\bibitem{CVE}
\BIBentryALTinterwordspacing
Mitre, ``\BIBforeignlanguage{en}{Common vulnerabilities and exposures}.''
  [Online]. Available: \url{https://cve.mitre.org/}
\BIBentrySTDinterwordspacing

\bibitem{cauligi2020constant}
S.~Cauligi, C.~Disselkoen, K.~v. Gleissenthall, D.~Tullsen, D.~Stefan, T.~Rezk,
  and G.~Barthe, ``Constant-time foundations for the new spectre era,'' in
  \emph{Proceedings of the 41st ACM SIGPLAN Conference on Programming Language
  Design and Implementation}, 2020, pp. 913--926.

\bibitem{lazardoes}
D.~Lazar, H.~Chen, X.~Wang, and N.~Zeldovich, ``Why does cryptographic software
  fail,'' in \emph{Proceedings of 5th Asia-Pacific Workshop on Systems}, p.~7.

\bibitem{carre2019cache}
S.~Carr{\'e}, A.~Facon, S.~Guilley, S.~Takarabt, A.~Schaub, and Y.~Souissi,
  ``Cache-timing attack detection and prevention: Application to crypto libs
  and pqc,'' in \emph{Constructive Side-Channel Analysis and Secure Design:
  10th International Workshop, COSADE 2019, Darmstadt, Germany, April 3--5,
  2019, Proceedings 10}.\hskip 1em plus 0.5em minus 0.4em\relax Springer, 2019,
  pp. 13--21.

\bibitem{reparaz2017dude}
O.~Reparaz, J.~Balasch, and I.~Verbauwhede, ``Dude, is my code constant time?''
  in \emph{Design, Automation \& Test in Europe Conference \& Exhibition
  (DATE), 2017}.\hskip 1em plus 0.5em minus 0.4em\relax IEEE, 2017, pp.
  1697--1702.

\bibitem{kocher1996timing}
P.~C. Kocher, ``Timing attacks on implementations of diffie-hellman, rsa, dss,
  and other systems,'' in \emph{Advances in Cryptology—CRYPTO’96: 16th
  Annual International Cryptology Conference Santa Barbara, California, USA
  August 18--22, 1996 Proceedings 16}.\hskip 1em plus 0.5em minus 0.4em\relax
  Springer, 1996, pp. 104--113.

\bibitem{brumley2011remote}
B.~B. Brumley and N.~Tuveri, ``Remote timing attacks are still practical,'' in
  \emph{Computer Security--ESORICS 2011: 16th European Symposium on Research in
  Computer Security, Leuven, Belgium, September 12-14, 2011. Proceedings
  16}.\hskip 1em plus 0.5em minus 0.4em\relax Springer, 2011, pp. 355--371.

\bibitem{bernstein2005cache}
D.~J. Bernstein, ``Cache-timing attacks on aes,'' 2005.

\bibitem{ge2018survey}
Q.~Ge, Y.~Yarom, D.~Cock, and G.~Heiser, ``A survey of microarchitectural
  timing attacks and countermeasures on contemporary hardware,'' \emph{Journal
  of Cryptographic Engineering}, vol.~8, pp. 1--27, 2018.

\bibitem{al2013lucky}
N.~J. Al~Fardan and K.~G. Paterson, ``Lucky thirteen: Breaking the tls and dtls
  record protocols,'' in \emph{2013 IEEE symposium on security and
  privacy}.\hskip 1em plus 0.5em minus 0.4em\relax IEEE, 2013, pp. 526--540.

\bibitem{genkin2015get}
D.~Genkin, I.~Pipman, and E.~Tromer, ``Get your hands off my laptop: physical
  side-channel key-extraction attacks on pcs: Extended version,'' \emph{Journal
  of Cryptographic Engineering}, vol.~5, pp. 95--112, 2015.

\bibitem{coron1999resistance}
J.-S. Coron, ``Resistance against differential power analysis for elliptic
  curve cryptosystems,'' in \emph{Cryptographic Hardware and Embedded Systems:
  First InternationalWorkshop, CHES’99 Worcester, MA, USA, August 12--13,
  1999 Proceedings 1}.\hskip 1em plus 0.5em minus 0.4em\relax Springer, 1999,
  pp. 292--302.

\bibitem{jancar2022they}
J.~Jancar, M.~Fourn{\'e}, D.~D.~A. Braga, M.~Sabt, P.~Schwabe, G.~Barthe, P.-A.
  Fouque, and Y.~Acar, ``“they’re not that hard to mitigate”: What
  cryptographic library developers think about timing attacks,'' in \emph{2022
  IEEE Symposium on Security and Privacy (SP)}.\hskip 1em plus 0.5em minus
  0.4em\relax IEEE, 2022, pp. 632--649.

\bibitem{NVD}
\BIBentryALTinterwordspacing
NIST, ``\BIBforeignlanguage{en}{National vulnerability database}.'' [Online].
  Available: \url{https://nvd.nist.gov/}
\BIBentrySTDinterwordspacing

\bibitem{Kholoosi2023}
\BIBentryALTinterwordspacing
M.~M. Kholoosi, A.~Babar, and cemal yilmaz, ``{Dataset for "Empirical Analysis
  of Software Vulnerabilities Causing Timing Side Channels"},'' 8 2023.
  [Online]. Available:
  \url{https://figshare.com/articles/dataset/Dataset_for_Empirical_Analysis_of_Software_Vulnerabilities_Causing_Timing_Side_Channels_/22725845}
\BIBentrySTDinterwordspacing

\bibitem{disselkoen2020finding}
C.~Disselkoen, S.~Cauligi, D.~Tullsen, and D.~Stefan, ``Finding and eliminating
  timing side-channels in crypto code with pitchfork.''\hskip 1em plus 0.5em
  minus 0.4em\relax TECHCON, 2020.

\bibitem{pereida2016make}
C.~Pereida~Garc{\'\i}a, B.~B. Brumley, and Y.~Yarom, ``Make sure dsa signing
  exponentiations really are constant-time,'' in \emph{Proceedings of the 2016
  ACM SIGSAC Conference on Computer and Communications Security}, 2016, pp.
  1639--1650.

\bibitem{barthe2018secure}
G.~Barthe, B.~Gr{\'e}goire, and V.~Laporte, ``Secure compilation of
  side-channel countermeasures: the case of cryptographic
  “constant-time”,'' in \emph{2018 IEEE 31st Computer Security Foundations
  Symposium (CSF)}.\hskip 1em plus 0.5em minus 0.4em\relax IEEE, 2018, pp.
  328--343.

\bibitem{pranckevivcius2017comparison}
T.~Pranckevi{\v{c}}ius and V.~Marcinkevi{\v{c}}ius, ``Comparison of naive
  bayes, random forest, decision tree, support vector machines, and logistic
  regression classifiers for text reviews classification,'' \emph{Baltic
  Journal of Modern Computing}, vol.~5, no.~2, p. 221, 2017.

\bibitem{rajput2020word}
N.~K. Rajput, B.~A. Grover, and V.~K. Rathi, ``Word frequency and sentiment
  analysis of twitter messages during coronavirus pandemic,'' \emph{arXiv
  preprint arXiv:2004.03925}, 2020.

\bibitem{bird2009natural}
S.~Bird, E.~Klein, and E.~Loper, \emph{Natural language processing with Python:
  analyzing text with the natural language toolkit}.\hskip 1em plus 0.5em minus
  0.4em\relax " O'Reilly Media, Inc.", 2009.

\bibitem{Lipp2018meltdown}
M.~Lipp, M.~Schwarz, D.~Gruss, T.~Prescher, W.~Haas, A.~Fogh, J.~Horn,
  S.~Mangard, P.~Kocher, D.~Genkin, Y.~Yarom, and M.~Hamburg, ``Meltdown:
  Reading kernel memory from user space,'' in \emph{27th {USENIX} Security
  Symposium ({USENIX} Security 18)}, 2018.

\bibitem{Kocher2018spectre}
P.~Kocher, J.~Horn, A.~Fogh, , D.~Genkin, D.~Gruss, W.~Haas, M.~Hamburg,
  M.~Lipp, S.~Mangard, T.~Prescher, M.~Schwarz, and Y.~Yarom, ``Spectre
  attacks: Exploiting speculative execution,'' in \emph{40th IEEE Symposium on
  Security and Privacy (S\&P'19)}, 2019.

\bibitem{aldaya2019port}
A.~C. Aldaya, B.~B. Brumley, S.~ul~Hassan, C.~P. Garc{\'\i}a, and N.~Tuveri,
  ``Port contention for fun and profit,'' in \emph{2019 IEEE Symposium on
  Security and Privacy (SP)}.\hskip 1em plus 0.5em minus 0.4em\relax IEEE,
  2019, pp. 870--887.

\bibitem{bao2022v}
L.~Bao, X.~Xia, A.~E. Hassan, and X.~Yang, ``V-szz: automatic identification of
  version ranges affected by cve vulnerabilities,'' in \emph{Proceedings of the
  44th International Conference on Software Engineering}, 2022, pp. 2352--2364.

\bibitem{croft2023data}
R.~Croft, M.~A. Babar, and M.~Kholoosi, ``Data quality for software
  vulnerability datasets,'' \emph{arXiv preprint arXiv:2301.05456}, 2023.

\bibitem{bernstein2019fast}
D.~J. Bernstein and B.-Y. Yang, ``Fast constant-time gcd computation and
  modular inversion,'' \emph{IACR Transactions on Cryptographic Hardware and
  Embedded Systems}, pp. 340--398, 2019.

\bibitem{java-github}
\BIBentryALTinterwordspacing
R.~Hauch, ``Use time constant algorithms when comparing passwords or keys,''
  2021. [Online]. Available:
  \url{https://github.com/apache/kafka/pull/10978/commits/b14a7a8cfcf44777771a622ccc919b11f4b65440}
\BIBentrySTDinterwordspacing

\bibitem{wordpress.org-documentation}
\BIBentryALTinterwordspacing
``Wordpress release notes-version 4.2.4,'' Feb 2019. [Online]. Available:
  \url{https://wordpress.org/documentation/wordpress-version/version-4-2-4/}
\BIBentrySTDinterwordspacing

\bibitem{php2-wordpress}
\BIBentryALTinterwordspacing
WordPress, ``Changeset 33536,'' 2015. [Online]. Available:
  \url{https://core.trac.wordpress.org/changeset/33536}
\BIBentrySTDinterwordspacing

\bibitem{C1-github}
\BIBentryALTinterwordspacing
G.~Doering, ``Use constant time memcmp when comparing hmacs in
  openvpn\_decrypt,'' 2013. [Online]. Available:
  \url{https://github.com/OpenVPN/openvpn/commit/f375aa67cc5e8c7e9639b8f020b192f948050eef}
\BIBentrySTDinterwordspacing

\bibitem{cpp2-github}
\BIBentryALTinterwordspacing
A.~Rojas, ``Added constant time comparison of jwt signatures.'' 2018. [Online].
  Available:
  \url{https://github.com/apache/mesos/commit/2c282f19755ea7518caf6f43e729524b1c6bdb23}
\BIBentrySTDinterwordspacing

\bibitem{enum1-github}
\BIBentryALTinterwordspacing
D.~Beck, ``[security-2566],'' 2022. [Online]. Available:
  \url{https://github.com/jenkinsci/jenkins/commit/957ef5902f2e40b6358e6d10f12b26f9dbd2f75a}
\BIBentrySTDinterwordspacing

\bibitem{enum3-github}
\BIBentryALTinterwordspacing
F.~Nägler, ``[security] prevent time based information disclosure,'' 2020.
  [Online]. Available:
  \url{https://github.com/TYPO3/typo3/commit/14929b98ecda0ce67329b0f25ca7c01ee85df574}
\BIBentrySTDinterwordspacing

\bibitem{croft2022noisy}
R.~Croft, M.~A. Babar, and H.~Chen, ``Noisy label learning for security
  defects,'' in \emph{Proceedings of the 19th International Conference on
  Mining Software Repositories}, 2022, pp. 435--447.

\bibitem{gousios2015work}
G.~Gousios, A.~Zaidman, M.-A. Storey, and A.~Van~Deursen, ``Work practices and
  challenges in pull-based development: The integrators perspective,'' in
  \emph{2015 IEEE/ACM 37th IEEE International Conference on Software
  Engineering}, vol.~1.\hskip 1em plus 0.5em minus 0.4em\relax IEEE, 2015, pp.
  358--368.

\end{thebibliography}
\end{document}